\RequirePackage[2020-02-02]{latexrelease}
\documentclass[a4paper]{revtex4}
\usepackage{inputenc}
\usepackage[english]{babel}

\usepackage{array,booktabs}
\usepackage{array} 
\usepackage{lipsum}   
\usepackage{calc}
\RequirePackage{graphicx}
\usepackage{color}
\usepackage{amsmath}
\usepackage{float}
\usepackage{calc}
\usepackage{pdflscape}
\usepackage{color}
\usepackage{float}
\usepackage{subfigure}
\usepackage[font=small,labelfont=bf]{caption}
\usepackage{graphicx}
\usepackage{pdflscape}
\usepackage{color}
\usepackage{float}
\usepackage{anyfontsize}
\usepackage[font=small,labelfont=bf]{caption}
\usepackage{graphicx}
\usepackage{amsmath}
\usepackage[nodisplayskipstretch]{setspace}
\renewcommand{\arraystretch}{1.5}
\usepackage{setspace}
\usepackage{tabularx}

\usepackage{float}
\usepackage{color}
\usepackage{amsmath}
\usepackage{float}
\usepackage{calc}
\usepackage{pdflscape}
\usepackage{xcolor}
\usepackage{float}
\usepackage{subfigure}
\usepackage[font=small,labelfont=bf]{caption}
\usepackage{graphicx}
\begin{document}
{\setlength\abovedisplayskip{4pt}}
\title{Annihilation of NMSSM neutralinos and Branching Ratios, Particle Decay Channel of lightest CP odd, even Higgs in NMSSM.}
\author{Gayatri Ghosh}
\email{gayatrighsh@gmail.com}
\affiliation{Department of Physics, Gauhati University, Jalukbari, Assam-781015, India}
\affiliation{Department of Physics, Cachar College, Silchar, Assam-788005, India}



\begin{abstract}                                                                                                      
The next$-$to$-$minimal supersymmetric standard model (NMSSM) featuring constrained mSUGRA model, has the capability to inherently anticipate a light dark matter component within the existing limitations encompassing Higgs data, sparticle$-$mass constraints, dark matter exploration, muon g-2. We examine neutralino dark matter within the NMSSM framework by conducting a comprehensive analysis of its parameter space. This involves evaluating neutralino capture and annihilation rates within the Sun. The exploration of potential detection strategies for neutralino dark matter in neutrino experiments hinges on the composition of neutralinos and their primary annihilation pathways. Our study also involves reassessing the maximum thresholds for branching ratios of lepton flavour violation decays $BR(\mu\rightarrow e+\gamma)$, $BR(\tau\rightarrow e+\gamma)$  by directly referencing the constrained limits on $ \Delta a_{\mu} $ from $ g_{\mu}-2 $ experiment. This work also presents constraints of muon flux, photon, positron and antiproton flux, specifically its independence from experimental intricacies and the universal applicability of recalculation coefficients across NMSSM model. In this work, we delve into the intricate calculations of muon flux resulting from the annihilation of dark matter, examining various sources including the cores of celestial bodies like the Sun and Earth, as well as cosmic diffuse neutrinos generated in dark matter annihilation within halos. A significant finding of our investigation lies in the distinct energy distribution of muons stemming from dark matter annihilation, contrasting sharply with those produced by atmospheric neutrinos. Furthermore, we meticulously consider the two scenarios of muon flux: upward flux, originating from muons created in the substratum beneath the detector, and contained flux, where muons are generated within the detector material itself. Additionally, we discuss the implications of neutrino flavor dependence and its ramifications for detection methodologies, providing valuable insights into the nuanced interplay between dark matter annihilation and neutrino observation. We also calculate the Branching Ratios, Particle Decay Channel of lightest CP odd, even Higgs in NMSSM. Within the scope of this research, we have chosen to utilize this NMSSM scenario as a case study to investigate the funnel$-$ annihilation mechanisms pertaining to light dark matter and the concealed Higgs decay. In this particular scenario, our findings reveal that there exist decay channel$-$annihilation mechanisms for the lightest supersymmetric particle $ \tilde{\chi_{1}^{0}} $, which include the $h_{2}, h_{1}, Z, W^{+}, W^{-}, G, s, S, b, B, c, C, a, A, d, D, l, L $ decay funnels. 	
                                                                                                          \end{abstract}

\maketitle
\section{Introduction}\label{Introduction}
\label{intro}                                                                                     The discovery of the 125 GeV Higgs boson at the Large Hadron Collider (LHC) in 2012, while validating the Standard Model (SM) at TeV energy scales, has also underscored the necessity for physics models beyond the SM to account for phenomena such as dark matter (DM). In this pursuit, various supersymmetric (SUSY) models {\color{red}\cite{GG}} offer the Lightest Supersymmetric Particle (LSP) and Next Lightest Supersymmetric Particle (NLSP) as a compelling candidate for the weakly interacting massive particle (WIMP), aligning with the natural predictions for dark matter, thus driving the exploration into new physics realms. Recent advancements in particle physics experiments have provided a wealth of information on SUSY models. With the data from Run-II Large Hadron Collider (LHC), scientists have been able to probe the characteristics of winos and higgsinos, reaching masses of approximately 1060 GeV for $ \tilde{m_{\chi_{1}^{0}}}\lesssim \text{400} GeV$ and 900 GeV for $ \tilde{m_{\chi_{1}^{0}}}\lesssim 400 \text{GeV}$, respectively. Here, $\chi_{1}^{0}$ represents the lightest neutralino, serving as the lightest supersymmetric particle (LSP) and a potential candidate for dark matter (DM) under R$-$parity conservation assumptions. Additionally, the LHC data have placed constraints on squarks with masses lighter than approximately 1850 GeV when the LSP is massless. Furthermore, the combined measurements of the muon anomalous magnetic moment by experiments at Fermilab and Brookhaven National Laboratory indicate a significant deviation from the Standard Model's prediction, hinting at potential new physics beyond the SM. While this deviation may be attributed to uncertainties in hadronic contributions, the possibility of SUSY effects has been widely discussed, promising insights into the mass spectra of electroweakinos and sleptons once confirmed. Moreover, the recent results from the LUX-ZEPLIN (LZ) experiment on direct DM search have set unprecedented limits on DM couplings to SM particles, further highlighting the relevance of SUSY in addressing fundamental questions in particle physics. These remarkable achievements motivate a comprehensive examination of their collective implications for SUSY theory.                                                                     
\par                                                                                                                  The minimal supersymmetric standard model (MSSM) stands out as one of the foremost contenders among new physics models, offering an elegant resolution to the hierarchy problem while designating the lightest neutralino as a viable candidate for dark matter (DM). However, challenges such as the µ-problem and the little hierarchy problem have surfaced, particularly under the scrutiny of recent LHC experiments. These issues find resolution in the next-to-minimal supersymmetric standard model (NMSSM) {\color{red}\cite{KN}}, which expands upon the Higgs sector of the MSSM by incorporating a gauge singlet field, denoted as $\hat{S}$. The development of a vacuum expectation value (VEV) for $\hat{S}$, represented by $ \upsilon s $, dynamically generates an effective $\mu-$term, naturally aligning its magnitude with the electroweak scale. Furthermore, interactions among Higgs fields $ \lambda \hat{S}\hat{H_{u}}\hat{H_{d}}$ in the NMSSM {\color{red}\cite{1,2,3}} contribute positively to the squared mass of the SM-like Higgs boson at the tree level. Additionally, the mass can undergo enhancement through singlet-doublet Higgs mixing, particularly if the Higgs boson represents the next to lightest CP even Higgs state. Consequently, the need for significant radiative corrections to the Higgs boson mass is mitigated, offering a solution to the little hierarchy problem within the framework of the NMSSM.                                                                                                                                                                                                                            \par	                                                                                             Dark matter (DM) stands as one of the most captivating enigmas within the realm of the Universe. Particle dark matter garners robust backing from copious evidence found in astrophysical observations {\color{red}\cite{G1}}. Amidst a plethora of proposed dark matter contenders postulated by theorists, the weakly interacting massive particle (WIMP), also referred to as thermal dark matter, emerges as a significant candidate. Its mass typically ranges from $O(1) MeV$ to $O(100) TeV$ {\color{red}\cite{G2}}, aligning with the natural interpretation of the DM relic density through the freeze$-$out mechanism. Moreover, it successfully elucidates the historical events of big bang nucleosynthesis and recombination during the early stages of the Universe. Nevertheless, prevailing negative outcomes from DM search endeavors, particularly those stemming from direct detection experiments, significantly diminish the feasible parameter space for WIMP at the electroweak scale. Consequently, this circumstance prompts a shift in focus towards lighter forms of DM, particularly exploring its annihilation mechanisms beyond the realm of direct scrutiny.                                                                               	                                                                                                    \par                                                                                                                Correlations between the Higgs boson and Dark Matter have been extensively examined for multiple decades, particularly following the initial indication of the 125 GeV Higgs boson {\color{red}\cite{aa}} in late 2011. In cases where the Higgs boson interacts with Dark Matter {\color{red}\cite{a,b,c,d,e,f,g,h,i,j,k}} and the mass of Dark Matter is less than half of the Higgs boson's mass, the unobservable decay of the Higgs boson provides an alternative method to indirectly detect Dark Matter. Recent analyses based on data from Run-I and Run-II at the Large Hadron Collider (LHC) have established that the upper limit for the invisible decay of the Higgs boson is 26 $\%$  as reported by ATLAS {\color{red}\cite{l}} and 19 $\%$ by CMS {\color{red}\cite{m}}. There exists considerable potential for investigating the characteristics of light Dark Matter originating from the invisible decay of the Higgs boson, a topic that is explored in the present study. The concept of Supersymmetry (SUSY) not only offers predictions for a natural Higgs boson resembling the Standard Model (SM) and a viable Dark Matter candidate, but also facilitates the unification of the three gauge interactions, thereby garnering significant interest. Numerous specific models and scenarios have been proposed within the framework of SUSY, with one notable scenario being the Next-to-Minimal Supersymmetric Standard Model (NMSSM) which can persist with the aforementioned advantages while adhering to current restrictions including Higgs boson data, bounds on sparticle masses, Dark Matter investigations, and the anomalous magnetic moment of the muon. This study focuses on examining the annihilation mechanisms of light Dark Matter arising from the invisible decay of the Higgs boson within the context of this scenario in the NMSSM.			                                                                                               \par                                                                                                                            Within the realm of SUSY, there exist various specific models and scenarios, one of which is a scenario within NMSSM with constrained mSUGRA model that can persist with the aforementioned advantages despite existing constraints such as Higgs data, sparticle-mass bounds, DM searches, and $ g_{\mu}-2$ {\color{red}\cite{GG1,GG2}}. Here we also explore the updated limits on $BR(\mu\rightarrow e+\gamma)$, $BR(\tau\rightarrow e+\gamma)$, SM Higgs boson mass, universal soft scalar masses $ m_{0} $, universal gaugino masses $ M_{1/2} $ and trilinear coupling from the constrainded limit on $\Delta a_{\mu}  $. Some studies on lepton flavour violation has also been done in {\color{red}\cite{GG, GG1,GG2}}. For the purposes of this study, we have chosen to focus on this particular scenario within the NMSSM in order to explore the mechanisms of annihilation for light DM stemming from the invisible decay of the Higgs.	                                                                                                                           \par	                                                                                                         Following this introduction, the subsequent section provides a summary of the pertinent characteristics of the NMSSM, along with an outline of the constraints utilized in our exploration of Chargino and Neutralino two body decays in our model. Moving forward to Section 3, we present the calculated Higgs like Branching ratio decays within the NMSSM. Finally, our findings and implications will be consolidated in Section 4, drawing conclusions based on the analyses conducted.																																																	\section{Theoretical foundations of the Next$-$to$-$Minimal Supersymmetric Standard Model (NMSSM), elucidating its key principles and fundamental concepts.}

                                                                                    					\begin{equation}
\textit{W}_{NMSSM} = \textit{W}_{YUKAWA} + \lambda \hat{S} \hat{H_{d}}\hat{H_{d}}+ \frac{\kappa}{3}\hat{s}	+ \mu \hat{H_{u}}.\hat{H_{d}} + \xi \hat{S}	+ \frac{1}{2}\mu^{'}\hat{S^{2}}																										
\end{equation}																																															where the Yukawa terms contained in $W_{Yukawa}$ resembles as those of the MSSM. $\hat{H_{u}}= (\hat{H_{u}}+,\hat{H_{u}0})^{T}$ and $\hat{H_{d}}= (\hat{H_{d}}0,\hat{H_{d}-})^{T}$ represents $SU(2)_{L}$ Higgs doublet. $ \lambda $, $ \kappa $ are dimensionless coupling parameters signifying invariant trilinear terms under $ Z_{3} $ symmetry. $ \mu $ and $ \mu^{'} $ are bilinear mass parameters and $ \xi $ is the singlet tadpole parameter which are written here explicitly to solve the cosmological domain wall problem. The $ \chi $ term may be erased by redefining $ \mu $	parameter. In this work the $ \chi $ term is set to zero. After the process of electroweak symmetry breaking, the interaction between the doublet and singlet scalar fields results in the creation of three charge-parity (CP)$-$even Higgs bosons denoted as $h_{1}, h_{2}, and h_{3}$, as well as two CP$-$odd Higgs bosons denoted as $a_{1} and a_{2}$. Additionally, the Higgsinos $\hat{H_{d}}\hat{H_{d}}$ (superpartners of the Higgs bosons) and the singlino $ \hat{S} $ mix with the gauginos  	$ \tilde{W^{0}}$ and $ \tilde{B} $ leading to the formation of five neutralinos $\chi_{i}^{0}$, $(i=1,2,3,4,5)$, where i ranges from 1 to 5. Henceforth, the symbol $\chi_{i}^{0}$  will represent the lightest neutralino, specifically the lightest supersymmetric particle (LSP) denoted as $\chi_{1}^{0}$, for the sake of simplicity.																\par 																									The NMSSM, also known as the NMSSM with nonuniversal Higgs masses (NUHM), postulates the convergence of gauginos and sfermions at the grand unified theory (GUT) scale, while excluding the Higgs sector from this unification [86–92]. Within the framework of the NMSSM, the gauginos are typically required to have substantial masses due to the upper limit on the gluino mass and the convergence of gauginos at the GUT scale. The strong interactions with the Standard Model (SM) sector result in the Higgsino-dominated Lightest Supersymmetric Particle (LSP) often leading to a minimal relic density. Consequently, in the NMSSM, the lightest neutralino  that predicts the correct relic density is predominantly singlino in nature. Should the singlino prove to be lighter than the SM$-$like Higgs, the scalar dominated by the singlet component is usually lighter than $125$ GeV; as a result, $h_{1}$ and $ a_{1} $ typically represent singlet$-$dominated scalars, while $ h_{2} $ usually corresponds to the SM$-$like Higgs with a mass around 125 GeV. By considering $\chi_{i}^{0}$ as a mixture of singlino and Higgsinos exclusively, it is feasible to express the interconnections between Higgs and $\chi$ through specific couplings $ \Gamma $ [93,94].													\begin{equation}
\Gamma_{h_{a}\chi\chi} = 1.414 \lambda(S_{a1}N_{12}N_{13} + S_{a2}N_{11}N_{13} + S_{a3}N_{11}N_{12})-1.414S_{a3}N_{13}^{2}
\end{equation}																							\begin{equation}
\Gamma_{a_{a}\chi\chi} = i[1.414 \lambda(P_{a1}N_{12}N_{13} + P_{a2}N_{11}N_{13} + P_{a3}N_{11}N_{12})-1.414_{a3}N_{13}^{2}],
\end{equation}	                                                                                                       The coefficients of $\hat{H_{u}}, \hat{h_{d}}$, and $\hat{S}$ in $ \chi $ are denoted as $N_{1i}$. In a similar manner, $S_{ai} and P_{ai}$ represent the coefficients of $\hat H_{u}$, $ \hat H_{d} $, and $S$ in $h_{a} and a_{a}$ correspondingly. In the case where singlino dominates $ \chi $, $ h_{2} $ exhibits characteristics similar to the Standard Model Higgs, while $h_{1}$ and $a_{1}$ are dominated by the singlet.		                                                                                                                              \begin{equation}
\Gamma_{h_{1}\chi\chi} = -1.414 \kappa, \Gamma_{a_{1}\chi\chi} = -i 1.414 \kappa,
\end{equation},	                                                                                                                    \begin{equation}
\Gamma_{h_{2}\chi\chi} = 1.414 \lambda N_{11} - 1.414 \kappa S_{23},
\end{equation}		                                                                                                     The invisible width of SM like Higgs $ h_{2} $ for 2$ m_{\chi} \leq m_{\phi}$ with $ \phi = h_{1}, h_{2}, a_{1} $																							\begin{equation}
\Upsilon _{\phi \rightarrow \chi \chi} = \frac{\Gamma^{2}\phi \chi \chi}{16 \pi} m_{\phi} (1-\frac{4m^{2}_{\chi}}{m^{2}_{\phi}})^{1.5}
\end{equation}																																																	\begin{equation}
\Upsilon _{\phi \rightarrow \chi \chi} = \frac{\Gamma^{2}\phi \chi \chi}{16 \pi} m_{\phi} (1-\frac{4m^{2}_{\chi}}{m^{2}_{\phi}})^{1.5}
\end{equation}																							In order to determine the relic density of the Lightest Supersymmetric Particle (LSP) denoted as $ \chi $ moving with a velocity $ \upsilon $, the initial step involves solving for the number density $n$ through the utilization of the Boltzmann equation.													\begin{equation}
\frac{dn}{dt} = -3H_{n} - <\upsilon \sigma_{eff}>(n^{2} -n^{2}_{eq}),
\end{equation}																							the Hubble rate denoted by H, the density in thermal equilibrium denoted by $ n_{eq} $, the total cross section of dark matter annihilation or coannihilation denoted by $ \sigma_{eff} $, and the thermal average of $ \upsilon \sigma_{eff} $ denoted by $ h \upsilon \sigma_{eff} $  are key parameters in the analysis. In the context of $s$-channel annihilation processes, a scenario unfolds where two dark matter particles undergo annihilation resulting in a medium boson $ \phi $, subsequently followed by the prompt decay of $ \phi $ into two Standard Model (SM) particles, a phenomenon known as $ \phi $-funnel annihilation. In accordance with perturbation theory, the cross section of $ \phi $-funnel annihilation leading to a fermion pair $ f\bar{f} $ can be explicitly expressed at the primary level.																			\begin{equation}
\sigma_{eff}^{\phi, f\bar{f}}= \frac{1}{2} \Gamma^{2}_{\phi \chi\chi} \frac{s\Upsilon_{\phi}\rightarrow f\bar{f}/m\phi}{(s-m_{\phi})^{2}+ m^{2}_{\phi}\Upsilon^{2}_{\phi}} \sqrt{1-\frac{4m^{2}_{\chi}}{s}}
\end{equation}																							the quantity $s$ represents the total center$-$of$-$mass energy squared, while $ \Upsilon_{\phi} $ and $ \Upsilon_{\phi\rightarrow f\bar{f}} $ denote the overall width and the $ f\bar{f} $ partial width of the decay of $\phi$, respectively. The comprehensive annihilation cross section $ \sigma_{eff} $ encompasses all channels, encompassing various particle mediums and final states. In the context of cold dark matter with nonrelativistic average velocities, it is possible to express $ \upsilon \sigma_{eff} $ as a series expansion in terms of $ \psi^{2} $.                                           																																																																																																							\begin{equation}
\upsilon \sigma_{eff} = a+ b\upsilon^{2} + \upsilon^{4}
\end{equation}																							The relic density can then be formulated as																\begin{equation}
\Omega_{\chi}h^{2} = m_{\chi}n_{0}\frac{h^{2}}{\rho_{c}}
\end{equation}																																																	\begin{equation}
\Omega_{\chi}h^{2} = \frac{m_{\chi}/T_{f}}{10}\sqrt{\frac{g_{*}}{100}}\frac{0.847 \times 10^{-27} cm^{3} s^{-1}}{\sigma_{eff} \upsilon}
\end{equation}																						Tthe current number density, denoted as $n_{0}$, is determined in consideration of the total effective degrees of freedom represented by $ g_{*} $, alongside the freezing-out temperature $ T_{f} $, and the critical density of the Universe denoted as $ \rho_{c} $, which is calculated as $ \rho_{c} = 3\frac{H_{0}^{2}}{8\pi G_{N}} $. The evaluation of the relic density of Dark Matter in Supersymmetric (SUSY) models is meticulously conducted through various publicly available software tools such as micrOMEGAs, where a comprehensive analysis of pertinent annihilation and coannihilation phenomena is incorporated.																																																																	\section{The Neutralino Sector}																			This mass matrix formulation matrix Neutralino provides a comprehensive understanding of the neutralino sector within  NMSSM, encompassing the bino field $\hat{B}$, the Wino field $\hat{W}$, Higgsino fields $ \hat{H_{u}^{0}} $, $ \hat{H_{d}^{0}} $, the singlino field $ \hat{S} $. Its structure, as delineated in Eq. (2) , elucidates the intricate interplay between these particles, shedding light on their masses and interactions crucial for the model's phenomenological implications. 																							\begin{equation}
M_{\tilde{\chi^{0}}} =  \begin{bmatrix}
M_{1} & 0 & -m_{Z}sin\theta_{W}cos\beta & m_{z} sin\theta_{W} & 0\\
& M_{2} &  m_{Z}cos\theta_{W}cos\beta & -m_{Z}cos\theta_{W} sin\beta & 0 \\
& & 0 & -\mu_{total} & -\frac{1}{2}\lambda_{\upsilon} sin\beta \\
& & & 0 & -\frac{1}{2}\lambda_{\upsilon}cos\beta \\
&&&&2\frac{\kappa}{\lambda}\mu_{eff} + \mu^{'}\\
\end{bmatrix}
\end{equation}																																																	$ M_{1} $, $ M_{2} $ are the gaugino masses. $ \mu_{total} = \mu_{eff} + \mu $	represents the Higssino masses. The neutralino mass eigen states are expressed by  									\begin{equation}
\chi_{i}^{0} = N_{i1}\psi_{1}^{0} + N_{i2}\psi_{2}^{0} + N_{i3}\psi_{3}^{0} + N_{i4}\psi_{4}^{0} + N_{i5}\psi_{5}^{0}
\end{equation}																							$ N_{i3} $, $ N_{i4} $ are the $ \hat{H_{d}^{^{0}}} $ and $ \hat{H_{u}^{0}} $ components in $ \chi_{i}^{0} $ respectively. $ N_{i5} $	represents the singlino components. Given the condition $ |m^{2}_{\tilde{\chi_{1}^{0}}} - \mu^{2}_{total}| > \lambda^{2}\upsilon^{2} $	and the presence of very highly massive gauginos, the singlino dominated mass $ \tilde{\chi_{1}^{0}} $ and its field composition are roughly estimated as 																	\begin{equation}
m_{\tilde{\chi_{1}^{0}}} =  \frac{2\kappa}{\lambda} \mu_{eff} + \mu^{'} + \frac{1}{2}\frac{\lambda^{2}\upsilon^{2}}{m^{2}_{\tilde{\chi_{1}^{0}}}-\mu^{2}_{total}}(m_{\tilde{\chi_{1}^{0}}}-\mu_{total}sin 2\beta), \hspace{0.2cm} N_{11} = 0, \hspace{0.2cm} N_{12} = 0
\end{equation}																							\begin{equation}
\frac{N_{13}}{N_{15}} = \frac{\lambda \upsilon}{\sqrt{2}\mu_{tot}} \frac{(m_{\tilde{\chi_{1}^{0}}}/\mu_{total})sin\beta - cos\beta}{1-(\frac{m_{\tilde{\chi_{1}^{0}}}}{\mu_{tot}})^{2}}, \hspace{0.2cm}	\frac{N_{14}}{N_{15}} = \frac{\lambda \upsilon}{\sqrt{2}\mu_{tot}} \frac{(m_{\tilde{\chi_{1}^{0}}}/\mu_{total})cos\beta - sin\beta}{1-(\frac{m_{\tilde{\chi_{1}^{0}}}}{\mu_{tot}})^{2}}		
\end{equation}																																																	\begin{equation}
N_{15}^{2} = \frac{[1-(\frac{m_{\tilde{\chi_{1}^{0}}}}{\mu_{total}})^{2}]^{2}}{[(\frac{m_{\tilde{\chi_{1}^{0}}}}{\mu_{total}})^{2}-2(\frac{m_{\tilde{\chi_{1}^{0}}}}{\mu_{total}})sin2\beta + 1](\frac{\lambda \upsilon}{\sqrt{2}\mu_{tot}})^{2} + [1-\frac{\lambda \upsilon}{\sqrt{2}\mu_{tot}}^{2}]^{2}}
\end{equation}																																																	The above expressions predicts many important characteristics of neutralino. Mass of neutralino depends upon the parameters $ \lambda. \kappa, tan\beta, \mu_{eff}, \mu_{total} $. $ \lambda $, $ \kappa $ are independent parameters in predicting neutralino mass. The field compositions in $ \chi_{1}^{0} $ are determined by $ tan\beta, \mu_{eff}, \lambda, m_{\chi_{1}^{0}}, \kappa $	in probing neutralino's properties.																																												\section{ The Higgs Sector}																				The Lagrangian of the soft breaking term of the Higgs field takes the following form:					\begin{equation}
-\textit{L}_{soft} = \lambda A_{\lambda}S H_{u}.H_{d} + \frac{1}{3}A_{\kappa}\kappa S^{3} + m_{3}^{2} H_{u}.H_{d} + \frac{1}{2}{m^{'}_{s}}^{2}S^{2} + h.c +  m^{2}_{H_{u}}|H_{u}|^{2} + m^{2}_{H_{d}}|H_{d}|^{2} + m^{2}_{S}|S|^{2}
\end{equation}																							$H_{u}$, $H_{d}$ and $S$ are the scalar components of multiplets of Higgs Superfields. Masses of $m^{2}_{H_{u}} $, $m^{2}_{H_{d}}$ and $m^{2}_{S}$ can be expressed in terms of Higgs vevs $ <H_{u}>^{0} = \frac{\upsilon_{u}}{\sqrt{2}}$, $ <H_{d}^{0}> = \frac{\upsilon_{d}}{\sqrt{2}}$ and $ <S> = \frac{\upsilon_{s}}{\sqrt{2}} $ by minimising the scalar potential. As one sees from the soft breaking Lagrangian the Higgs sector in NMSSM model is defined by the yukawa couplings, $ \lambda, \kappa, \mu_{eff} $ and the trilinear couplings $ A_{\lambda}, A_{\kappa} $, the bilinear soft mass parameters, $ \mu, \mu^{'} $. related to soft breaking parameters $ m_{3}^{2}, m_{S}^{2} $.  The elements of CP$ - $	even Higgs boson mass matrix $ M_{S}^{2} $ are as follows:																																	\begin{equation}
M^{2}_{S,11} = \frac{2[\mu_{eff}(\lambda A_{\lambda + \kappa \mu_{eff}} + \lambda \mu^{'})] + \lambda m_{3}^{2}}{\lambda sin 2 \beta} + \frac{1}{2}(2m^{2}_{Z} - \lambda^{2}\upsilon^{2})sin^{2}2\beta, 
\end{equation}																							\begin{equation}
M^{2}_{S,12} =  -\frac{1}{4}(2m^{2}_{Z} - \lambda^{2}\upsilon^{2})sin 4\beta, 
\end{equation}																							\begin{equation}
M^{2}_{S,13} = -\frac{1}{\sqrt{2}}(\lambda A_{\lambda + 2\kappa \mu_{eff}} + \lambda \mu^{'})\upsilon cos2\beta, 
\end{equation}																											\begin{equation}
M^{2}_{S,22} = m^{2}_{Z}cos^{2} 2 \beta + \frac{1}{2}\lambda^{2}\upsilon^{2}sin^{2}2\beta
\end{equation}																							\begin{equation}
M^{2}_{S,23} = \frac{\upsilon}{\sqrt{2}}[2\lambda (\mu_{eff + \mu})-(\lambda A_{\lambda + 2 \kappa\mu_{eff}}+ \lambda \mu^{'}) Sin2\beta]
\end{equation}																																																	\begin{equation}
M^{2}_{S,33} = \frac{\lambda(A_{\lambda + \mu^{'}})sin 2 \beta}{4\mu_{eff}}\lambda \upsilon^{2} + \frac{\mu_{eff}}{\lambda}(\kappa A_{\kappa} + \frac{4\kappa^{2}\mu_{eff}}{\lambda} + 3 \kappa \mu^{'}) - \frac{\mu}{2 \mu_{eff} \lambda^{2}\upsilon^{2}}.
\end{equation}																							And the CP odd Higgs fields are 																		\begin{equation}
M^{2}_{P,11} =  \frac{2[\mu_{eff}(\lambda A_{\lambda} + \kappa \mu_{eff} + \lambda \mu^{'}) + \lambda m_{3}^{2}]}{\lambda Sin 2\beta},
\end{equation}																							\begin{equation}
M^{2}_{P,22} =  \frac{(\lambda A_{\lambda} + \kappa \mu_{eff} + \lambda \mu^{'})sin 2\beta}{4 \mu_{eff}}\lambda \upsilon^{2} -\frac{\kappa \mu_{eff}(3 A_{\kappa} + \mu^{'})}{\lambda} - \frac{\mu}{2 \mu_{eff}}\lambda^{2}\upsilon^{2} - 2m^{'2}_{S},
\end{equation}																																																	\begin{equation}
M^{2}_{P,12} =  \frac{\upsilon}{\sqrt{2}}(\lambda A_{\lambda} -2\kappa \mu_{eff} -\lambda \mu^{'})
\end{equation}																							$ h_{i} = \lbrace h,H,h_{s} \rbrace $	and $a_{i} = \lbrace A_{H}, A_{S} \rbrace $ the mass eigen states leads to the following:															\begin{equation}
h_{i} = V_{h_{i}}^{NSM} H_{NSM} + V_{h_{i}}^{SM}H_{SM} + V_{h_{i}}^{S}Re[S],
\end{equation}																							\begin{equation}
a_{i} = V_{P,a_{i}}^{NSM} A_{NSM} + V^{S}_{P,a_{i}} Im[S]
\end{equation}																																																	These mass eigen states are obtained by unitary rotation matrices $ V $ and $ V_{P} $ to diagonalise $ M_{S}^{2} $ and $ M_{P}^{2} $ respectively. $ h $ scalar is the SM like Higgs boson discovered at LHC. $ H $, $ A_{H} $ represents the heavy doublet$ - $	dominated states. $ h_{s} $, $ A_{S} $ are the singlet dominated states. The charged Higgs states $ H^{\pm} $ takes the form: 						\begin{equation}
m^{2}_{H^{\pm}} = m^{2}_{A} + m^{2}_{W} - \lambda^{2}\upsilon^{2},
\end{equation}
\begin{equation}																					H^{\pm} = m^{2}_{A} + m^{2}_{W} - \lambda^{2}\upsilon^{2}
\end{equation}																							where $ H^{\pm} = cos\beta H_{u}^{\pm} + sin\beta H_{d}^{\pm}$. Here $ H^{\pm} = cos\beta H_{u}^{\pm} + sin\beta H_{d}^{\pm}$ and, $ m^{2}_{A} = 2\frac{[\mu_{eff}(\lambda A_{\lambda} + \kappa \mu_{eff}) + \lambda \mu^{'}) + \lambda m_{3}^{2}]}{\lambda sin 2\beta}$. Furthermore, extensive searches for additional Higgs bosons, including $H, A_{H}, h_{s}, A_{s}, and H^{\pm}$, have been conducted at the LHC with considerable intensity, as documented in references.							Asymptotically for $ \lambda\longrightarrow 0 $, the mass matrix elements becomes,  					\begin{equation}
M^{2}_{S,11} = m^{2}_{A} + m^{2}_{Z}sin^{2}2\beta,\hspace{0.4cm} M^{2}_{S,12} = \frac{-1}{2} m^{2}_{Z}sin 4\beta \hspace{0.4cm} M^{2}S_{1,3} = 0
\end{equation}																							\begin{equation}
M^{2}_{S,22} = m^{2}_{Z}cos^{2}2\beta, \hspace{0.1cm}M^{2}_{S,23} = 0,\hspace{0.1cm} M^{2}_{S,33} = \frac{\mu_{eff}}{\lambda}(\kappa A_{\kappa} + \frac{4\kappa^{2}\mu_{eff}}{\lambda} + 3\kappa\mu^{'}), 
\end{equation}																							\begin{equation}
M^{2}_{P,11} = m^{2}_{A} \hspace{0.1cm} M^{2}_{P,22} = -\frac{\kappa\mu_{eff}}{\lambda}(3A_{\kappa}+\mu^{'})-2m^{'2}_{S} \hspace{0.1cm} M^{2}_{P,12} = 0
\end{equation}																							Thus it is found that the masses of the heavy doublet$-$dominated scalars are primarily dictated by the parameters $A_{\lambda}$ and $m_{3}$. Parameters $ A_{\kappa} $ and $ m_{s}^{'} $ exists in $M^{2}_{S,33}$ and $ M^{2}_{P,22} $. Thus $ m_{h_{s}} $, $ m_{A_{s}} $ varies with $ A_{\kappa} $ and $ m_{s^{'}} $.m.,\textit{l} Here, The mass range of $m_{H^{\pm}}$ spans from approximately 1050 GeV to 5000 GeV, aligning well with the constraints observed in the LHC's search for $H^{\pm}$ particles.
\section{Muon g-2}
Another aspect of the SUSY source of the muon $g-2$, denoted as $a_{\mu}^{SUSY}$, involves loops mediated by a smuon coupled with a neutralino, as well as those featuring a muon$-$type sneutrino interacting with a chargino. The expression of $a_{\mu}^{SUSY}$ in the mass insertion approximation [32], aiming to elucidate its fundamental characteristics are presented here. In this approximation's lowest order, the contributions to $a_{\mu}^{SUSY}$ can be classified into four categories: WHL, BHL, BHR, and BLR, where $W, B, H, L,$ and$ R$ denote the wino, bino, higgsino, and left$-$handed and right$-$handed smuon fields, respectively. These contributions stem from Feynman diagrams involving transitions such as $\tilde{W} - \tilde{H_{d}}$, $\mu_{L}-\mu_{R}$, and they exhibit the following mathematical forms:
\section{Calculation}                                                                                                                    
In our investigation, the muon flux originating from muon neutrinos resulting from dark matter (DM) annihilation hinges on several factors, including the flux of muon neutrinos calculated using Eq. (35), as well as attenuation, tau neutrino regeneration, and neutrino mixing during transit to the detector. While neutrino mixing and tau neutrino regeneration may influence muon neutrino flux in scenarios of DM annihilation within the Sun, teir impact is deemed negligible for our analysis due to the specific range of mass of dark mattter  values considered, where only moderate modifications to the flux are anticipated. We thus omit these effects from our calculations. Particularly for upward events where muons are generated outside the detector, the energy loss of muons assumes significance. To simplify our evaluation, we initially focus on contained events, wherein muon energy loss can be more straightforwardly addressed. Subsequently, we turn our attention to the neutrino-induced flux of muons, whether produced within the detector or in its vicinity, while accounting for muon energy loss. To model the detector's response, we employ an effective area, which, in the case of IceCube, can be parameterized as a function of muon energy at the detector, providing a robust framework for our analysis. The flux of neutrinos of flavor i from dark matter annihilation into standard model particles is as follows:
\begin{equation}
\frac{d\phi_{\nu}}{dE_{\nu}} = \frac{\tau_{A}}{4 \pi R^{2}} \Sigma B_{F}(\frac{dN_{\nu}}{dE_{\nu}})_{F,i}
\end{equation}
where $ \frac{dN_{\nu}}{dE_{\nu}} $ is the differential energy spectrum of
neutrino flavor i from production of particles in channel F.
Neutrino detectors, such as IceCube/DeepCore, are designed with the objective of identifying the capture of Lightest Supersymmetric Particles (LSPs) within the Sun through the detection of neutrinos generated during the LSP annihilation process. The primary observational signal is characterized by the detection of neutrino-induced muons traversing the detector. In this analysis, we investigate the flux of neutrinos and neutrino-induced muons within the $tan \beta = 10$ and $55$ planes, as well as across the WMAP-preferred focus-point, coannihilation, and funnel-region strips (the latter applicable to $tan \beta = 55$ only). Additionally, we explore the neutrino flux spectra. The determination of the neutrino spectrum within a detector entails two main stages: (1) the generation of neutrinos from LSP annihilations, and (2) the transmission of these neutrinos from the Sun's core to the detector. Generally, low-energy neutralinos do not directly undergo annihilation into neutrinos; instead, neutrinos are produced during decays or showers of the primary annihilation particles like W bosons, top quarks, or tau leptons (e.g., $ \chi\chi  \rightarrow \tau \bar{\tau}$, with $\tau \rightarrow \mu \bar{\nu_{\mu}}\nu_{\tau} $). The assessment of neutrino spectra within such showers is intricate due to their occurrence in the dense solar core, leading to potential energy loss by the primary particles before decay. Following neutrino production, they traverse through the Sun where they can experience charged-current or neutral-current interactions that either absorb the neutrinos or diminish their energies, respectively. While the Sun is generally permeable to neutrinos below 100 GeV, it becomes opaque as energies reach 200$-$300 GeV, resulting in significant suppression of high-energy neutrinos. The consideration of neutrino oscillations between different species during their passage from the Sun to Earth is imperative. Our analysis relies on the neutrino and neutrino-induced muon spectra derived from WimpSim (utilized within DarkSUSY), which simulate both production and propagation processes to generate spectra for various LSP annihilation channels. Furthermore, neutrino production and propagation have been simulated in a prior study. For our current neutrino spectrum analysis, we exclude Higgs annihilation channels and solely include annihilations into quarks, leptons, W and Z bosons.  \begin{figure}[h]
\centering
\includegraphics[width=0.6\linewidth]{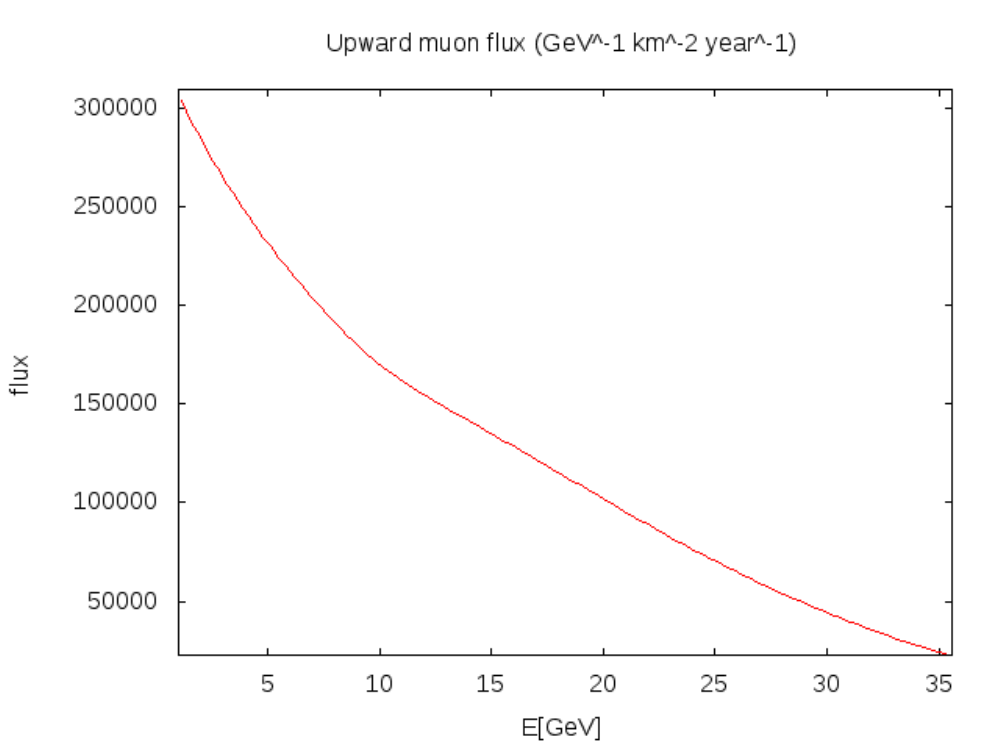}
\caption{Upward Muon fluxes obtained from dark matter annihilation in the halos producing cosmic diffuse neutrinos.}
\label{fig:flavour}
\end{figure}	
High-energy muons generated in neutrino charged-current interactions undergo energy loss as they traverse through the intervening medium, whether it be rock or ice. The process of energy dissipation is crucial to understanding the characteristics of muons detected at the surface. The average energy loss experienced by muons with an initial energy $E$ over a short distance 
$dz$ through a medium with a density $ \rho $ can be described by the Bethe-Bloch formula. This formula provides an estimate of the energy loss per unit path length for charged particles traversing a material. Mathematically, the average energy loss $ \frac{-dE}{dz}$ is given by			\begin{equation}
-\frac{-dE}{dz} = Kz^{2}\frac{Z}{A}\frac{1}{\beta^{2}}[ln(\frac{2 m_{e}c^{2}\beta^{2}\gamma^{2}T_{max}}{I^{2}})-\beta^{2}-\delta(\beta_{\gamma})]
\end{equation}
K is the constant of integration, $ z $ is the charge on the incident particle. $Z$ is the atomic number of the medium. A is the atomic mass of the medium. $ \beta $ is the velocity of muon divided by speed of light. $ \gamma $ is the Lorentz factor. $ T_{max} $ is the maximum kinetic energy transferred to the electrons. $ I $ is the mean excitation energy of the medium. $ \delta $ is the density corrected term.

This equation encapsulates the intricate interplay between the properties of the incident muon, the characteristics of the traversed medium, and the underlying atomic structure of the material. By integrating this formula along the muon's trajectory, we can ascertain the total energy loss incurred by the muon as it traverses through the medium, providing essential insights into the expected energy spectrum of muons reaching the detector. This equation encapsulates the intricate interplay between the properties of the incident muon, the characteristics of the traversed medium, and the underlying atomic structure of the material. By integrating this formula along the muon's trajectory, we can ascertain the total energy loss incurred by the muon as it traverses through the medium, providing essential insights into the expected energy spectrum of muons reaching the detector.

\par 
Here we seek for weak interactions which are transmitted by intermediate bosons, called $W$ , of a mass $ m_{W} $ which is not much bigger than 1 GeV. Since we are only interested in very high energy neutrinos (from about 10 GeV to several hundred GeV) the dominant reactions are the coherent processes

\begin{equation}
\nu_{\mu} + Z \rightarrow \mu^{-} + W^{+} +z
\end{equation}

\begin{equation}
\bar{\nu_{\mu}} + z \rightarrow \mu^{+} \longrightarrow \mu^{+} + W^{-} +Z
\end{equation}

Where Z is the target nucleus. This reaction takes place at a vertical depth, say, (h cos 0) from the surface of the earth and produce a muon of energy $ \mu $ moving at an angle $ \theta $ with respect to the upward vertical direction. Figure 1 shows muon flux in units of $ year^{-1} $, $ km^{-2} $, integrated over $ 2 \pi $ solid angle and muon energy from 5 GeV up. Only the muons produced by reactions (37) and (38) are included. 
\par 
The term contained muon flux refers to the flux of muons that are entirely the muons produced by neutrino interactions within a detector. The contained muon flux would then represent the number of muons that are produced and remain within the detector volume, as opposed to those that pass through or interact outside the detector. In particle physics or astrophysics, experiments often measure the flux of muons to study various phenomena, such as cosmic ray interactions, neutrino oscillations, or high-energy astrophysical processes. Figure 2 represents contained muon flux in units of $ year^{-1} $, $ km^{-2} $, integrated over $ 2 \pi $ solid angle and muon energy from 5 GeV up. 

The neutrino flux which is produced in the atmosphere can be decomposed into three terms:
\begin{equation}
N_{\nu} = N_{\nu} (\mu) + N_{\nu} (\tau) + N_{\nu} (K)
\end{equation}
due to the decays of $ \mu $, $ \tau $ and $ K $ respectively. By using the observed muon spectrum in cosmic rays, $D(E_{\mu}, cos \theta)dE_{\mu}d\omega$ of muon decays for muon Energy $ E_{\mu} $ which decay in the atmosphere per unit area per unit time can be calculated.  The following equation gives the neutrino flux.                                  

\begin{equation}
N_{\nu} (\mu) = \frac{1}{3} \int D(E_{\mu}, cos \theta)[5-9 (\frac{\nu}{E_{\mu}}^{2} + 4 (\frac{\nu}{E_{\mu}}^{3})]\frac{dE_{\mu}}{\mu}
\end{equation}
In Eq.(40) we include only the $ \mu- $neutrinos from $ \mu $-decays, since e-neutrinos can not directly produce $ \mu^{\pm} $. The neutrino flux due to pion decays are as follows:
\begin{equation}
N_{\nu}(\pi) = \frac{7}{3}\int f(E_{\pi})[\frac{B_{\pi}}{B_{\pi} + E_{\pi}Cos \theta}] \frac{dE_{pi}}{E_{\pi}}
\end{equation}
where $ f(E) = 0.16 (\frac{E}{GeV})^{-2.6} $ per $GeV-cm^{2}-sterad$ and $ B_{\pi} = 120 GeV $. To determine the neutrino flux resulting from $K-$decays, we assume that the energy distribution of $K-$ mesons mirrors that of $\pi-$mesons. Consequently, $N_{\nu}(K)$ can be expressed as
\begin{equation}
N_{\nu} (K) = (b r) \int f(E_{K})[\frac{B_{K}}{B_{K} + E_{K} cos \theta}](\frac{dE_{K}}{E_{K}})
\end{equation}
Where $ B_{K} = 900$ GeV. $ r $ is the average production ratio of $ \frac{K}{\pi} $ at the same energy in a high energy nucleon-nucleon collision, $ b $ is the branching ratio of $ K_{\mu 2} $ decay. Here $ b r = 10 $ percent. Due to the significantly higher fractional energy transferred to the neutrino in a $K_{\mu 2}-$decay compared to a $ \pi_{\mu 2} $-decay, $N_{\nu}(K)$ plays a crucial role in the production of high-energy neutrinos. On the other hand, the observed muon spectrum remains largely unaffected by $ K_{\mu 2} $-decay, assuming the branching ratio (b r) is minimal. In Fig. 3 we present neutrino flux in units of $(year)^{-1} GeV^{-3} (Kilo Meter)^{-1}$ , integrated over $2\pi$ solid angle and muon energy from 5 GeV up to 70 GeV. Only the neutrino fluxes  produced by Eq.[39-42] are considered. In Fig. 4 and Fig. 5 we present positron flux and anti proton flux in units of $ GeV^{-1} (cm)^{-2}sr^{-1} cm^{-2}$ , integrated over $2\pi$ solid angle and  energy from 5 GeV up to 70 GeV. In particle physics and cosmology, relic abundances versus temperature describe how the quantities of certain particles, typically those from the early universe, change as the universe cools over time. These relic particles include dark matter candidates, neutrinos, and other exotic particles that were produced in the early moments after the Big Bang. In the early universe, temperatures were extremely high, and particles were in thermal equilibrium. This means that particles and antiparticles were constantly being created and annihilated. As the universe expanded, it cooled. When the temperature dropped below a certain threshold, some particles could no longer maintain equilibrium with the rest of the particle soup. This decoupling happens when the interaction rates of particles become slower than the expansion rate of the universe. For particles like dark matter candidates, as the universe cools, they "freeze out" of thermal equilibrium. This means that their number densities become effectively constant because the annihilation rate drops below the expansion rate. The temperature at which this happens is crucial for determining their relic abundance. The relic abundance is the remaining density of these particles after freeze-out. It depends on the particles' interactions, masses, and the specific dynamics of the early universe. For dark matter, the relic abundance is a key parameter for models predicting the amount of dark matter we observe today. The relationship between relic abundances and temperature can be depicted in a graph where the x-axis represents temperature (often in units of MeV or GeV), and the y-axis represents the abundance of the relic particle (often normalized to the total number of particles or the critical density of the universe). in Fig. 6 we illustrate the relationship between relic abundances and temperature. In Fig. 7, Fig. 8 and Fig. 9 we present photon flux and positron flux in units of $ GeV^{-1} (cm)^{-2}sr^{-1} cm^{-2}$ , integrated over $2\pi$ solid angle and  energy from 5 GeV up to 70 GeV for 124.54 SM like Higgs boson.
\begin{figure}[h]
\centering
\includegraphics[width=0.6\linewidth]{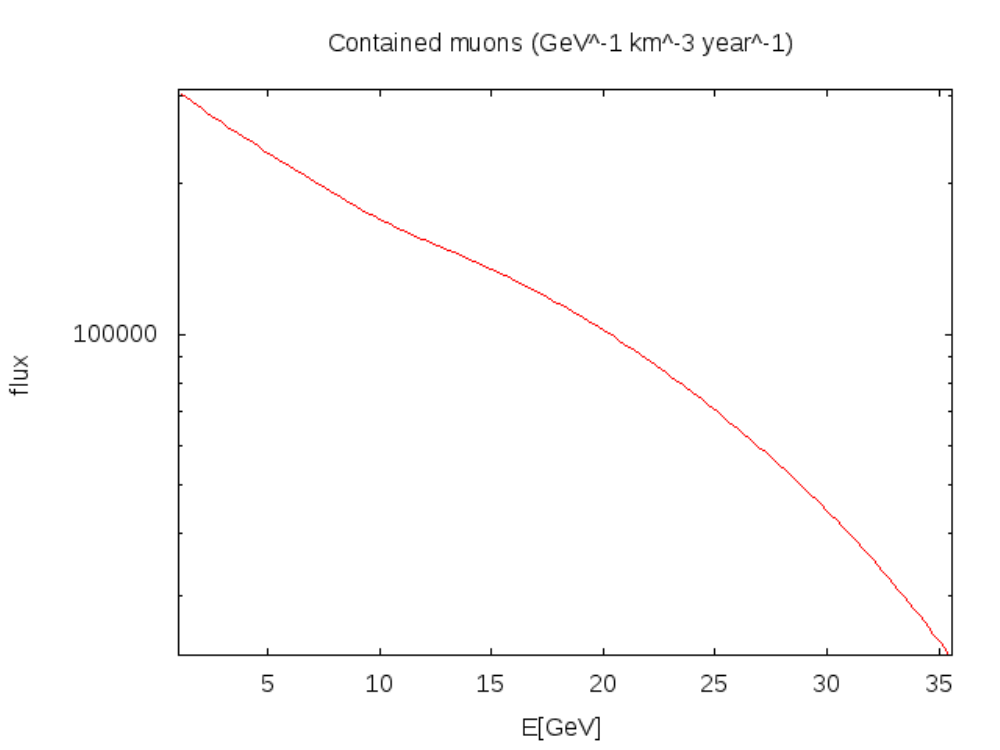}
\caption{Energy spectrum of contained muons produced by cosmic ray neutrinos interacting with the earth's crust.}
\label{fig:flavour}
\end{figure}																								\begin{figure}[h]
\centering
\includegraphics[width=0.6\linewidth]{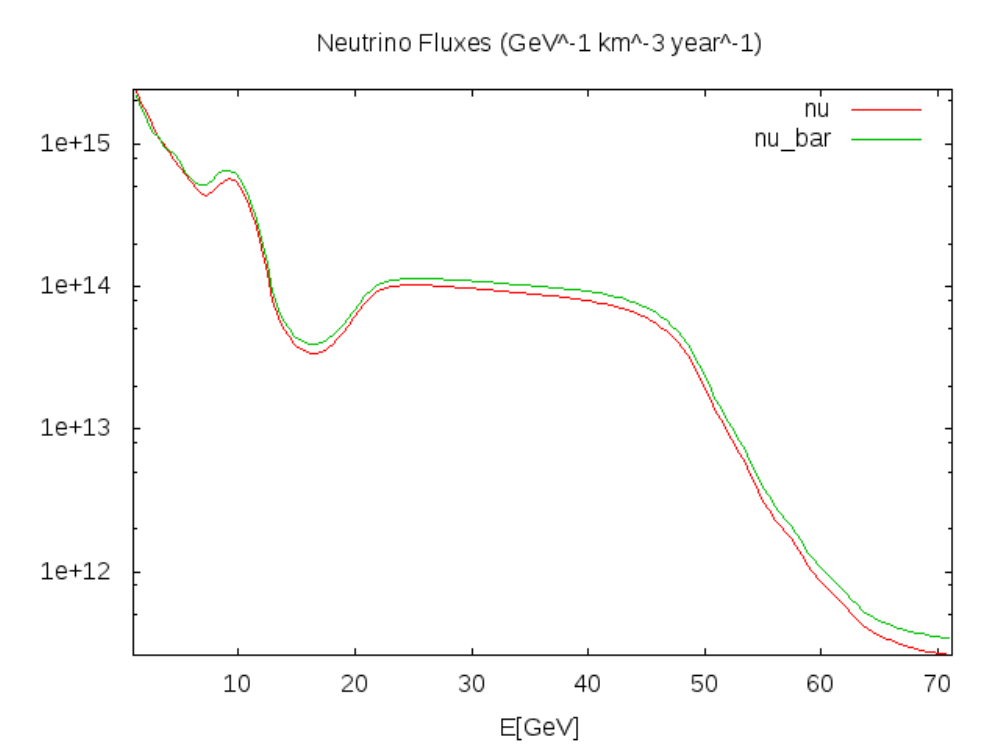}
\caption{Value of Neutrino flux as a function of Energy for calculated values of relic density in this work as detected by Neutrino Telescope.}
\label{fig:flavour}
\end{figure}																																																	\begin{figure}[h]
\centering
\includegraphics[width=0.6\linewidth]{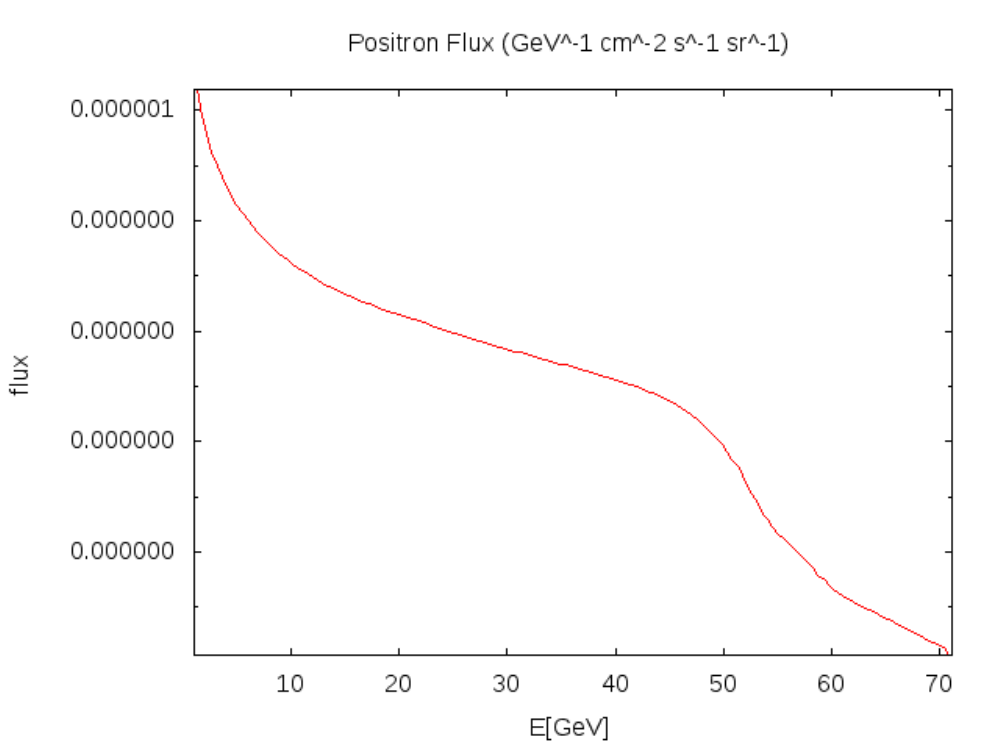}
\caption{Value of Positron flux as a function of Energy for calculated values of relic density in this work.}
\label{fig:flavour}
\end{figure}																																																	\begin{figure}[h]
\centering
\includegraphics[width=0.6\linewidth]{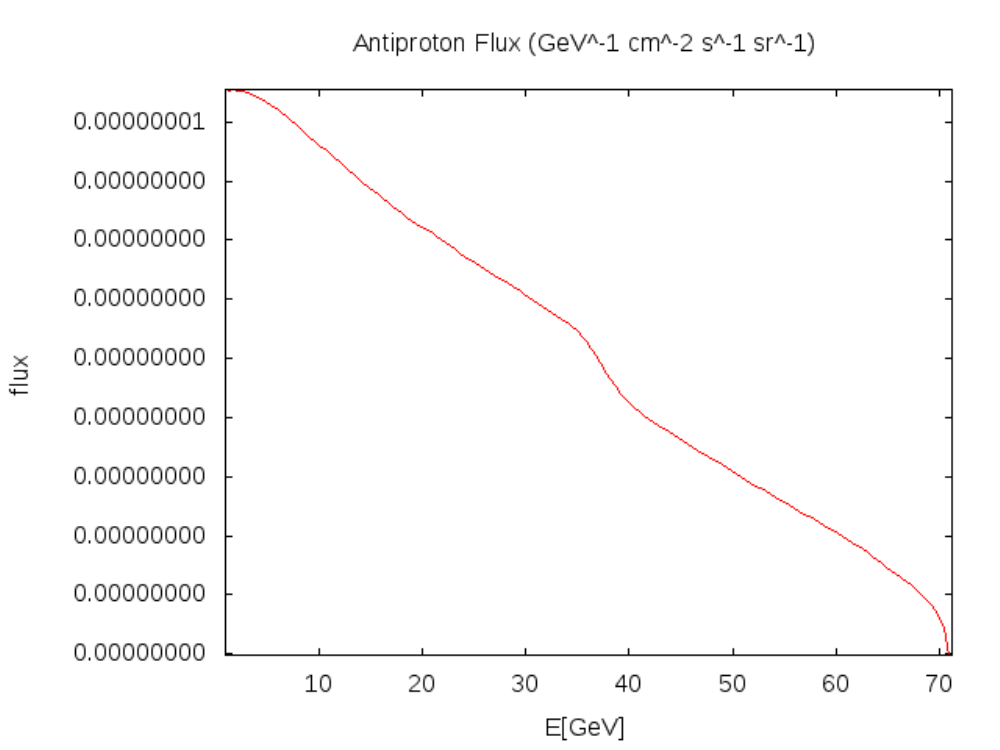}
\caption{Value of AntiProton flux as a function of Energy for calculated values of relic density in this work.}
\label{fig:flavour}
\end{figure}																																																	\begin{figure}[h]
\centering
\includegraphics[width=0.6\linewidth]{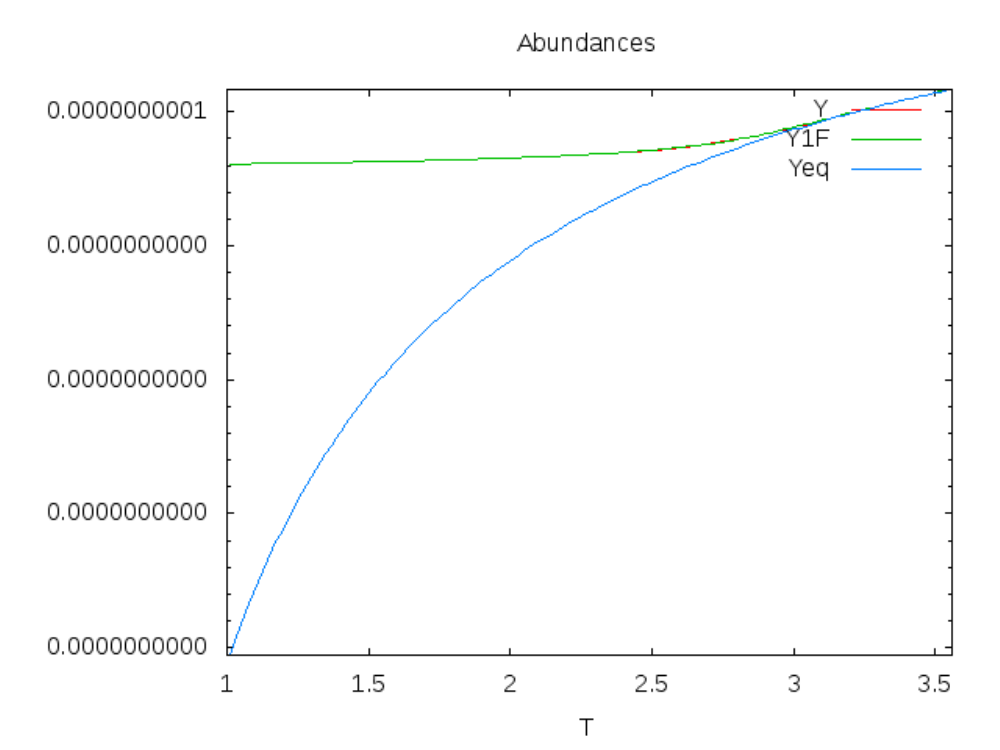}
\caption{Value of $ Y $ as a function of Temperature for calculated values of relic density in this work. This figure illustrates the evolution of neutralino dark matter abundances in the early universe as a function of temperature. The green line represents the equilibrium abundance, showing how neutralino are in thermal equilibrium at high temperatures. The freeze-out temperature, indicates how particles decouple from the thermal bath and stop interacting significantly. The solid line represents the relic abundance after freeze-out, which remains constant as the universe continues to cool. This constant abundance corresponds to the number density of relic particles observed in the present-day universe.}
\label{fig:flavour}
\end{figure}																																																	\begin{figure}[h]
\centering
\includegraphics[width=0.6\linewidth]{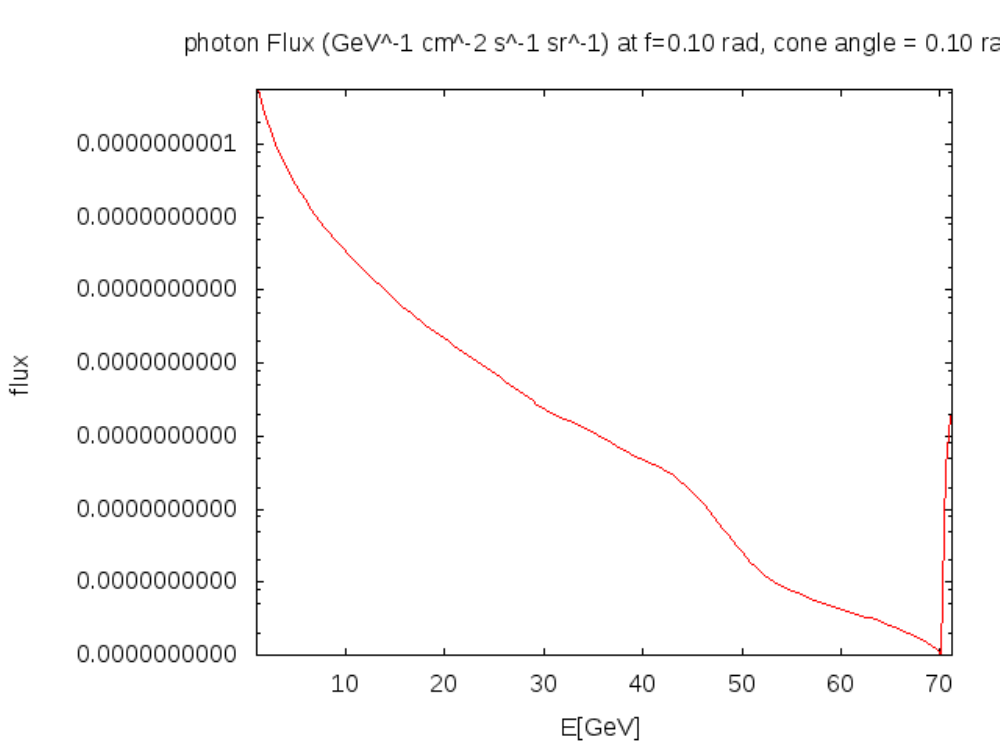}
\caption{Value of Photon flux as a function of Energy for calculated values of relic density in this work for 124.54 SM like Higgs boson.}
\label{fig:flavour}
\end{figure}                                                                                                                             																											\begin{figure}[h]
\centering
\includegraphics[width=0.6\linewidth]{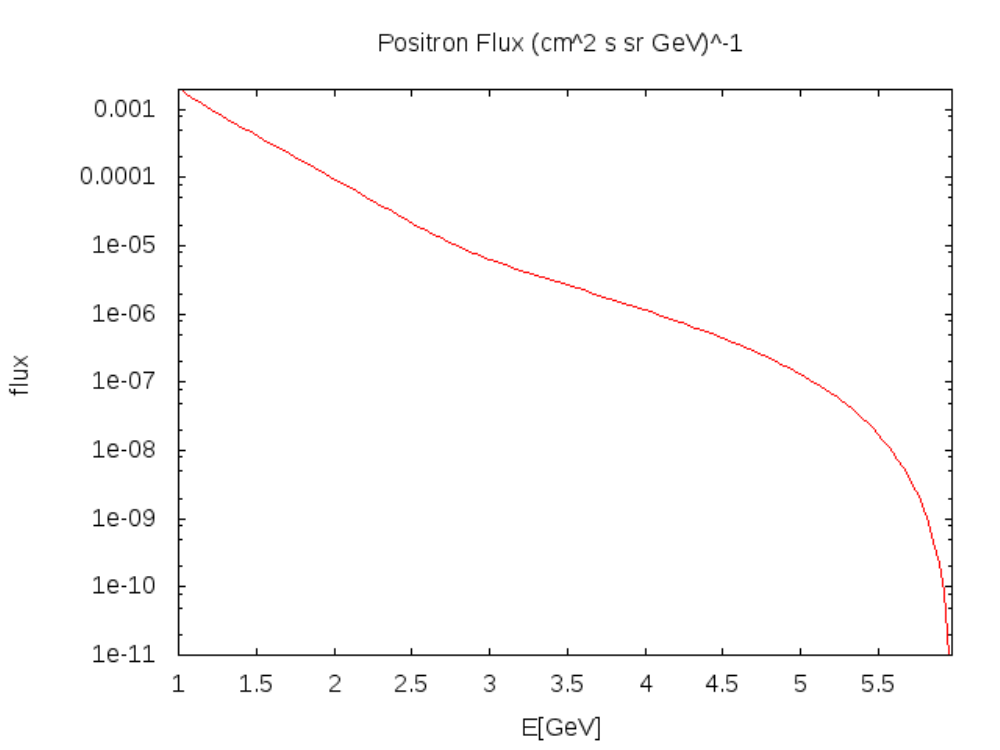}
\caption{Value of Positron flux as a function of Energy for calculated values of relic density in this work for 124.54 SM like Higgs boson.}
\label{fig:flavour}
\end{figure}                                                                                                                             				                                                                                                       \begin{figure}[h]
\centering
\includegraphics[width=0.6\linewidth]{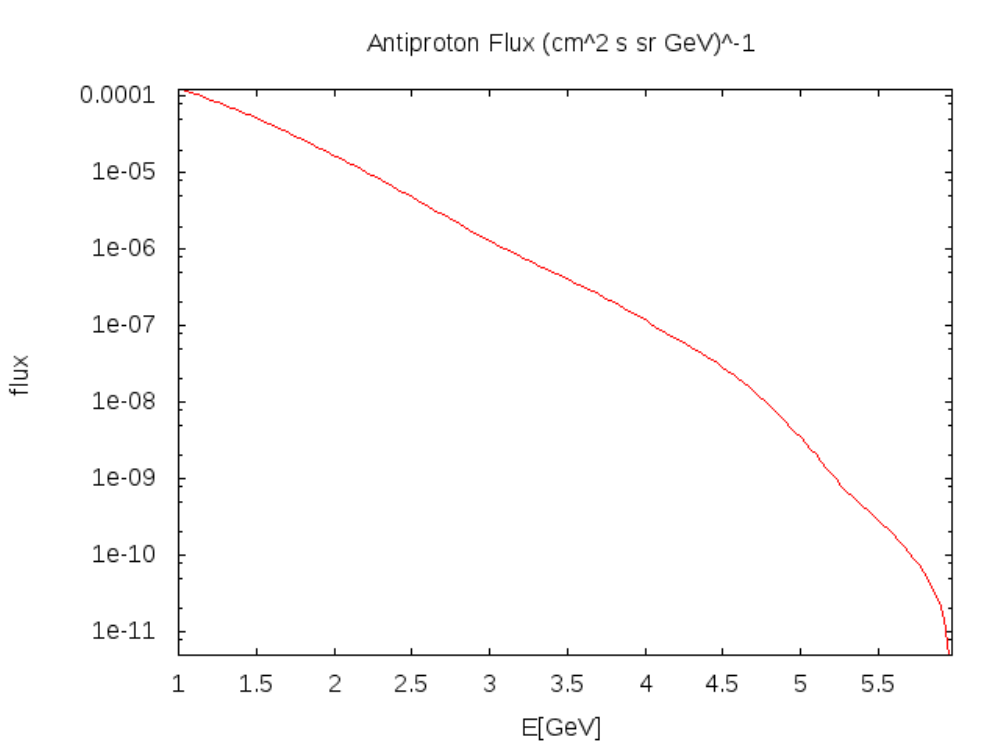}
\caption{Value of Amtiproton flux as a function of Energy for calculated values of relic density in this work for 124.54 SM like Higgs boson..}
\label{fig:flavour}
\end{figure}                                                                                                                             				                                                                                                     																										\begin{table}[ht]
\renewcommand{\arraystretch}{1.5}
\begin{center}
\begin{tabular}{|c|c|c|}
\hline 
Higgs & Higgs Masses &  Widths\\ 
\hline 
  $h_{1}$ &$18.05$ \hspace{0.1cm} GeV & $2.80E-04$\\
\hline
   $h_{2}$ &$122.00$ \hspace{0.1cm} GeV & $1.91E-02$\\
\hline
 $h_{3}$ &$505.06$ \hspace{0.1cm} GeV & $4.94E+00$\\  
\hline
$h_{a}$ &$248.54$ \hspace{0.1cm} GeV & $7.97E-01$\\  
\hline
$h_{b}$ &$499.53$ \hspace{0.1cm} GeV & $5.97E+00$\\  
\hline
$H^{+}$ &$492.81$ \hspace{0.1cm} GeV & $5.24E+00$\\  
\hline
 
\end{tabular}
\end{center}
\caption{Masses of Higgs and Decay Width.}
\end{table}
The decay width of the Higgs boson refers to the probability of the Higgs decaying into various particles. The total decay width is inversely related to the particle's lifetime. The decay width provides crucial information about the Higgs' interactions with other particles. The total decay width of the Higgs boson is very small, approximately 4.07 MeV (million electron volts). This corresponds to a relatively long lifetime for a particle of its mass. Here we have calculated the  T several decay channels of SM like Higgs boson, each with a different branching ratio (probability of occurring) which are presented in Table I, Table II, Table III. The Next-to-Minimal Supersymmetric Standard Model (NMSSM) is an extension of the Minimal Supersymmetric Standard Model (MSSM) that includes an additional gauge singlet superfield. This extension results in a richer Higgs sector with a total of seven Higgs bosons: three neutral CP even Higgs bosons, two neutral CP odd Higgs bosons, and two charged Higgs bosons. Here, we shall focus on the six neutral Higgs bosons in Table. III. $ h_{1} $ is the lightest CP even Higgs boson. This particle is often considered to be similar to the Standard Model (SM) Higgs boson but can mix with other CP even states, altering its properties. $ h_{2} $ is the second lightest CP$-$ even Higgs boson. Its mass and couplings can vary significantly depending on the parameters of the NMSSM. It could be heavier or lighter than the SM-like Higgs, depending on the model specifics. $ h_{3} $ is the heaviest CP$-$even Higgs boson in the NMSSM. This state is typically more massive and may have smaller couplings to the SM particles, making it harder to detect. $ h_{a} $ is the lightest CP odd Higgs boson. In the NMSSM, this particle can be very light, even less than half the mass of the SM like Higgs boson, leading to potential new decay channels. . $ h_{b} $ is the heavier CP odd Higgs boson. This state is usually more massive and can mix with the singlet component introduced in the NMSSM. $ H^{+} $ is the heavier CP$-$odd Higgs boson. This state is usually more massive and can mix with the singlet component introduced in the NMSSM.

\begin{table}[ht]
\renewcommand{\arraystretch}{1.5}
\begin{center}
\begin{tabular}{|c|c|}
\hline 
 Branching Ratio &  Particle Decay  Channel\\ 
\hline 
  $9.101241E-01$ & $h_{1}\rightarrow b, B$\\
\hline
 $7.971977E-02$ & $h_{1}\rightarrow l, L$\\
\hline
    $9.693996E-03$ & $h_{1}\rightarrow G, G$\\ 
\hline
 $1.990615E-04$ & $h_{1}\rightarrow d, D$\\              
\hline
 $1.990615E-04$ & $h_{1}\rightarrow s, S$\\ 
\hline
 $6.042197E-05$ & $h_{1}\rightarrow c, C$\\ 
\hline
$3.513639E-06$ & $h_{1}\rightarrow a, A$\\ 
\hline
$2.225840E-07$ & $h_{1}\rightarrow u, U$\\  
  \hline 
\end{tabular}
\end{center}
\caption{Branching Ratio, Particle  Decay channel of lightest CP odd Higgs $ h_{1} $ of total width $ 2.801915 E-04$ .}
\end{table}																								                                                                                                                                  \begin{table}[ht]
\renewcommand{\arraystretch}{1.5}
\begin{center}
\begin{tabular}{|c|c|}
\hline 
 Branching Ratio &  Particle Decay  Channel\\ 
\hline 
  $8.046774E-01$ & $h_{2}\rightarrow h_{1}, h_{1}$\\
\hline
 $1.236143E-01$ & $h_{2}\rightarrow b, B$\\
\hline
    $3.292189E-02$ & $h_{2}\rightarrow W^{+}, W^{-}$\\ 
\hline
 $1.539206E-02$ & $h_{2}\rightarrow G, G$\\              
\hline
 $1.356124E-02$ & $h_{2}\rightarrow l, L$\\ 
\hline
 $5.3561257E-03$ & $h_{2}\rightarrow c, C$\\ 
\hline
$3.912415E-03$ & $h_{2}\rightarrow Z, Z$\\ 
\hline
$4.688780E-04$ & $h_{2}\rightarrow A, A$\\  
  \hline                                                                                                                                                               
$3.224592E-05$ & $h_{2}\rightarrow d, D$\\                                                                                                                                                                \hline                                                                                            $3.224592E-05$ & $h_{2}\rightarrow s, S$\\                                                             \hline                                                                                                 $3.129216E-05$ & $h_{2}\rightarrow u, U$\\                                                             \hline                                                                                                       \hline\end{tabular}
\end{center}
\caption{Branching Ratio, Particle  Decay channel of lightest CP even Higgs $ h_{2} $ of total width $ 1.919972 E-02$ .}
\end{table}																								                                                                                                                                  \begin{table}[ht]
\renewcommand{\arraystretch}{1.5}
\begin{center}
\begin{tabular}{|c|c|}
\hline 
 Annihilation Cross Section & Annihilation Decay Channel of Neutralinos\\ 
\hline 
  $9.03E-01$ & $o1 o1\rightarrow b, B$\\
\hline
 $7.55E-02$ & $o1 o1\rightarrow l, L$\\
\hline
    $2.08E-02$ & $o1 o1\rightarrow G, G$\\ 
\hline
 $1.75E-04$ & $o1 o1 \rightarrow d, D$\\              
\hline
 $1.75E-04$ & $o1 o1 \rightarrow s, S$\\ 
\hline
 $6.042197E-05$ & $o1 o1 \rightarrow c, C$\\ 
\hline
$3.513639E-06$ & $o1 o1 \rightarrow a, A$\\ 
\hline
$2.225840E-07$ & $o1 o1 \rightarrow u, U$\\  
  \hline 
\end{tabular}
\end{center}
\caption{Indirect detection of neutralino of total annihilation cross section $3.96 E-25$ $cm^{3}/sec$ of 122 GeV SM like Higgs Boson.}
\end{table}                                             		 	                                                                                                              	                                                                                                               \begin{table}[ht]
\renewcommand{\arraystretch}{1.5}
\begin{center}
\begin{tabular}{|c|c|}
\hline 
 Annihilation Cross Section & Annihilation Decay Channel of Neutralinos\\ 
\hline 
  $9.81E-01$ & $o1 o1 \rightarrow Z, h_{1}$\\
\hline
 $1.04E-02$ & $o1 o1 \rightarrow W^{+}, W^{-}$\\
\hline
    $6.32E-03$ & $o1 o1 \rightarrow b, B$\\ 
\hline
 $7.81E-04$ & $o1 o1 \rightarrow c, C$\\              
\hline
 $7.12E-04$ & $o1 o1 \rightarrow l, L$\\ 
\hline
 $6.24E-04$ & $o1 o1 \rightarrow Z,Z$\\ 
\hline
$1.12E-04$ & $o1 o1 \rightarrow h_{1}, h_{2}$\\ 
\hline
$1.02E-04$ & $o1 o1 \rightarrow G, G$\\  
  \hline 
\end{tabular}
\end{center}
\caption{Indirect detection of neutralino of total annihilation cross section $3.96 E-25$ $cm^{3}/sec$ of 124.54 GeV SM like Higgs Boson.}
\end{table}	                                                                                                  \begin{figure}[h!]																							\centering																																													\includegraphics[width=5cm]{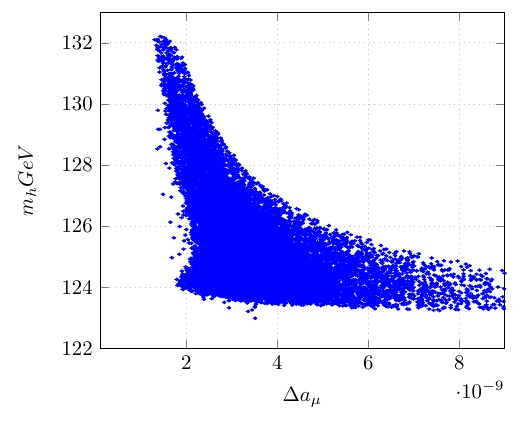}	                                                                    \includegraphics[width=5cm]{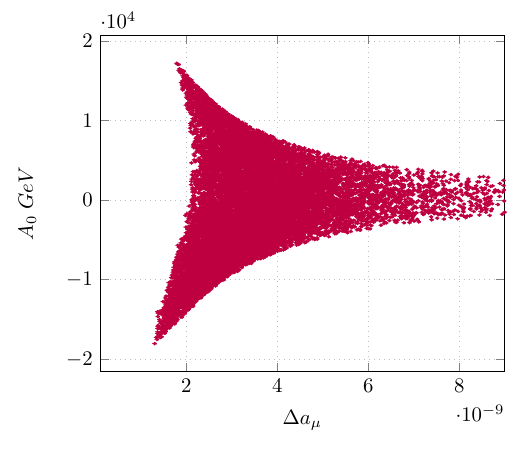}														\caption{Mass of second lightest CP even Higgs boson $m_{h_{2}}$ and trilinear coupling $ A_{0} $  as a function of $\Delta a_{\mu}  $ in our NMSSM model satisfying the current constraint on LFV decays BR($ \mu\rightarrow e +\gamma $) in the left and right panel respectively.}																								 \end{figure}																									
\begin{figure}[h!]																							\centering																																													\includegraphics[width=5cm]{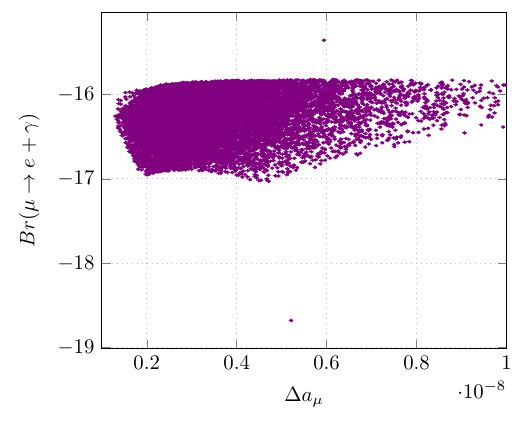}	                                                                    \includegraphics[width=5cm]{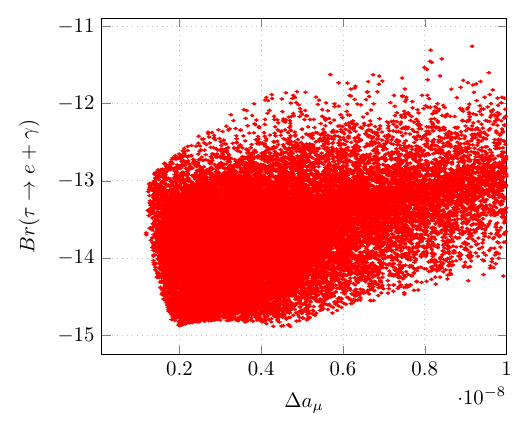}														\caption{Log 10 (BR($ \mu \rightarrow e +\gamma $)), Log 10 (BR($ \tau \rightarrow e +\gamma $) as a function of  $\Delta a_{\mu}  $ in our constrained NMSSM model in the left and right panel respectively.}																								 \end{figure} 																										 																										 		                                                                                                      \begin{figure}[h!]																							\centering																																													\includegraphics[width=5cm]{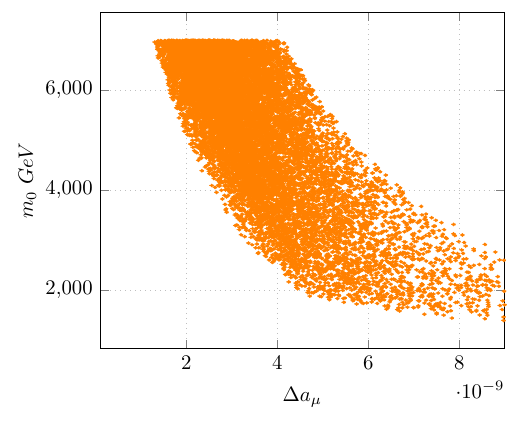}	                                                                    \includegraphics[width=5cm]{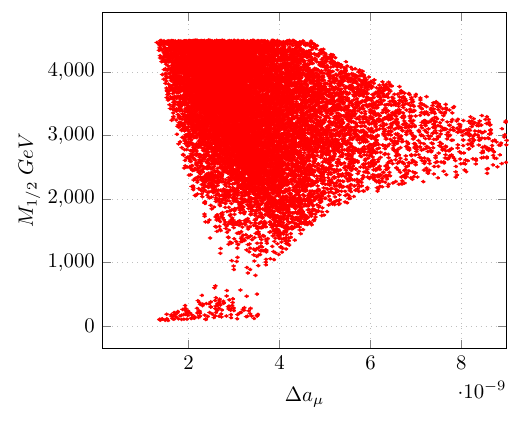}														\caption{Mass of soft scalar masses $m_{0}$ and soft gaugino masses $ M_{1/2} $  as a function of  $\Delta a_{\mu}  $ in our model satisfying the current constraint on LFV decays BR($ \mu \rightarrow e +\gamma $), BR($ \tau \rightarrow e +\gamma $) and  BR($ \tau \rightarrow \mu +\gamma $)  in the left and right panel respectively.}													\end{figure}																							The current experimental value of the anomalous magnetic moment of the muon,  $\Delta a_{\mu}  $ 
, is based on precise measurements and comparisons with theoretical predictions. The anomalous magnetic moment of the muon, $ \Delta a_{\mu}  = \frac{g-2}{2} $. $g$ is the gyromagnetic ratio, has been measured with high precision, and any deviation from the Standard Model prediction is denoted as $ \Delta a_{\mu}$. The difference between the experimental value and the Standard Model prediction is 
\begin{equation}
\Delta a_{\mu} = a_{\mu}^{exp} - a_{\mu}^{SM} = (251 \pm 59) \times 10^{-11}
\end{equation} 
This discrepancy of about 4.2 sigma (as of the latest combined results) suggests a potential hint of new physics beyond the Standard Model. The experimental value primarily comes from measurements by the E821 experiment at Brookhaven National Laboratory and the more recent results from the Muon g-2 experiment at Fermilab.	In Fig. 10 in the context of the Next to Minimal Supersymmetric Standard Model (NMSSM), the mass of the second lightest CP even Higgs boson $ m_{h_{2}} $ and the trilinear coupling $A_{0}$ are explored as functions of the anomalous magnetic moment of the muon, $\Delta a_{\mu}$. In our NMSSM model, these relationships are examined while ensuring compliance with the current constraints on lepton flavor violating (LFV) decays, specifically the branching ratio of $\mu \rightarrow e + \gamma$. In the left panel, the dependence of $ m_{h_{2}}$ on $\Delta a_{\mu}$ is analyzed. This exploration helps to understand how variations in the muon's magnetic moment, possibly indicating new physics, influence the mass spectrum of the extended Higgs sector in the NMSSM. A consistent $\Delta a_{\mu}$ with observed deviations from the Standard Model predictions can provide insight into viable ranges for $m_{h_{2}}$. In the right panel, the trilinear coupling $ A_{0}$, a parameter crucial for determining the scalar potential and thus the properties of the Higgs bosons, is plotted against $\Delta a_{\mu}$. Ensuring the branching ratio BR($\mu \rightarrow e + \gamma $) is within experimental limits imposes significant constraints on $ A_{0}$, as LFV processes are highly sensitive to supersymmetric contributions. These panels collectively demonstrate the interplay between Higgs sector parameters, the muon's anomalous magnetic moment, and stringent LFV constraints, highlighting how $\Delta a_{\mu}$ can shape viable NMSSM parameter space while adhering to current experimental observations. As is evident from Fig. 10 the experimental constraints on $ \Delta a_{\mu} = (2.51 \pm 0.59) \times 10^{-11} $	implies mass of $ m_{h_{2}} $ around 124.54 GeV which is the exact calculated value of the mass of the second lightest CP even Higgs boson $ m_{h_{2}} $ in our work. 
 In Fig. 11 we plot Log 10  BR($\mu \rightarrow e + \gamma $), Log 10  BR($\tau \rightarrow e + \gamma $) as a function of  $\Delta a_{\mu}  $ in our constrained NMSSM model in the left and right panel respectively. We find that the explored soft susy parameter space in our NMSSM model satisfies the current experimental limit on $ BR( \mu \rightarrow e +\gamma )\leq 4.2 \times 10^{-13} $, $ BR( \tau \rightarrow e +\gamma )\leq 4.3 \times 10^{-8} $ as constrained by Eq. (43). In Fi. 12 we present the mass of soft scalar masses $m_{0}$ and soft gaugino masses $ M_{1/2} $  as a function of  $\Delta a_{\mu}  $ in our model satisfying the current constraint on LFV decays BR($ \mu \rightarrow e +\gamma $), BR($ \tau \rightarrow e +\gamma $) and  BR($ \tau \rightarrow \mu +\gamma $)  in the left and right panel respectively. It is found from the figure that heavy soft scalar masses around   
5800-6500 GeV are allowed in our model set by the current experimental constraint on $ \Delta a_{\mu} = a_{\mu}^{exp} - a_{\mu}^{SM} = (251 \pm 59) \times 10^{-11} $. Gaugino masses, $ M_{\frac{1}{2}} $ lying in the range $ 2000-4500  $ GeV are allowed which is consistent with $ \Delta a_{\mu} = a_{\mu}^{exp} - a_{\mu}^{SM} = (251 \pm 59) \times 10^{-11} $. 
\section {Conclusion}
This work pertains to the potential of flux of neutrino, muon, photon and positron flux within the NMSSM framework.  A comprehensive scan across the NMSSM parameter space was conducted utilizing the NMSSMTools package. In our model, the lightest neutralino, predominantly bino, primarily annihilates into $bB$ pairs. Nonetheless, annihilation into $W^{+}W^{-}$ and $tt^{-}$ pairs also holds promise for detection at neutrino telescopes. For the general NMSSM model, a notable competition was found between $b\bar{b}$ and $\tau^{+}\tau^{-}$ annihilation channels for neutralino masses below the$ W-boson$ mass. Conversely, $W^{+}W^{-}$ and $t\bar{t}$ channels become comparable to annihilation into scalar pairs for heavier neutralinos. Examination revealed that NMSSM neutralinos with masses around $200 GeV$ interact with nucleons in the Sun mainly through axial interactions. As the mass increases, scalar interactions play a more significant role, leading to predominantly spin-independent scattering for masses exceeding $500 GeV$. The composition of annihilation products influences the fractions of low (soft) and high (hard) energy spectra of emitted particles, including neutrinos. The likelihood of a neutrino telescope detecting neutrinos from projected neutralino models is influenced by the energy threshold of the detector for neutrino-induced upward-going muons. The main distinguishable feature of this work lies in the detection of the mass of the second lightest CP even Higgs boson $ m_{h_{2}} $. As is evident from Fig. 10, the experimental constraints on $\Delta a_{\mu} = (251 \pm 59) \times 10^{-11}$ imply a mass of $ m_{h_{2}} $ around 124.54 GeV, which is the exact calculated value of the mass of the second lightest CP-even Higgs boson $ m_{h_{2}}$ in our work. This precise alignment underscores the robustness of our model and its compatibility with the latest experimental data. In summary, the novelty of this work lies in its integrative approach, detailed parameter space analysis, and the application of rigorous constraints from both $\Delta a_{\mu}$ and LFV decays. These contributions may provide valuable insights and constraints for the NMSSM model, potentially guiding future research and experiments in the quest for new physics.
	                                                                                                                    \section{Acknowledgement}
GG would like to thank would like to thank  University Grants Commission RUSA, MHRD, Government of India for financial support to carry out this work. She would also like to thank Department of Physics Cachar College, Silchar, Assam, India in this regard.                                                                  


\begin{thebibliography}{}
\bibitem{GG}  Kalpana Bora, Gayatri Ghosh, \textit{Charged Lepton Flavor Violation $\mu \rightarrow e \gamma$ in $ \mu $-$ \tau $ Symmetric SUSY SO(10) mSUGRA, NUHM, NUGM and NUSM Theories and LHC}, {\color{blue}Eur. Phys. J. \textbf{C75}, 9, 428, (2015)}.

\bibitem{KN} Higgsino Asymmetry and Direct-Detection Constraints of Light Dark Matter in the NMSSM with Non-Universal Higgs Masses, Kun Wang, Jingya Zhu, Quanlin Jie, 2020, Chin.Phys.C 45 (2021) 4, 041003 • e-Print: 2011.12848 [hep-ph]; Funnel annihilations of light dark matter and the invisible decay of the Higgs boson, Kun Wang Jingya Zhu, Phys.Rev.D 101 (2020) 9, 095028, arXiv: 2003.01662 [hep-ph]; A Novel Scenario in the Semi-constrained NMSSM, Kun Wang, Jingya Zhu(Wuhan U.), JHEP 06 (2020) 078; 2002.05554 
\bibitem{1} "Neutralino annihilation in the NMSSM with light singlet-like Higgs bosons, Belanger, G., et al.                                                     
                                                                                                                                    \bibitem{2}                                                                                           Branching ratios for Higgs boson decays into neutralinos and charginos in the NMSSM , Li, T., et al.                                                                                                       \bibitem{3} NMSSM Higgs Boson Decay to Neutralino Pairs at the LHC , Han, C., et al. 	
\bibitem{G1}  G. Bertone and D. Hooper, Rev. Mod. Phys. 90, 045002 (2018).
\bibitem{G2}  T. Hambye and M. H. G. Tytgat, Phys. Lett. B 683, 39
(2010) 
\bibitem{aa} J. Cao, Y. He, L. Shang, W. Su, and Y. Zhang, J. High
Energy Phys. 03 (2016) 207                                                                                  
\bibitem{a} G. Belanger, F. Boudjema, A. Cottrant, R. M. Godbole,
and A. Semenov, Phys. Lett. B 519, 93 (2001).
\bibitem{b} M. C. Bento, O. Bertolami, and R. Rosenfeld, Phys. Lett. B
518, 276 (2001).
\bibitem{c} K. Belotsky, D. Fargion, M. Khlopov, R. Konoplich, and
K. Shibaev, Phys. Rev. D 68, 054027 (2003).
\bibitem{d} R. M. Godbole, M. Guchait, K. Mazumdar, S. Moretti, and
D. P. Roy, Phys. Lett. B 571, 184 (2003).
\bibitem{e} H. Davoudiasl, T. Han, and H. E. Logan, Phys. Rev. D 71,
115007 (2005).
\bibitem{f} E. Accomando et al., https://doi.org/10.5170/CERN-2006-
009 (2006).
\bibitem{g} P. Draper, T. Liu, C. E. M. Wagner, L.-T. Wang, and H.
Zhang, Phys. Rev. Lett. 106, 121805 (2011).
\bibitem{h} Y. Cai, X.-G. He, and B. Ren, Phys. Rev. D 83, 083524
(2011).
\bibitem{i} J.-J. Cao, K.-i. Hikasa, W. Wang, and J. M. Yang, Phys.
Lett. B 703, 292 (2011).
\bibitem{j} X.-G. He and J. Tandean, Phys. Rev. D 84, 075018 (2011).
\bibitem{k} M. Pospelov and A. Ritz, Phys. Rev. D 84, 113001 (2011).                                                                                     
 \bibitem {l} ] M. Aaboud et al. (ATLAS Collaboration), Phys. Rev. Lett.
122, 231801 (2019).                                                                                    
\bibitem{m}  A. M. Sirunyan et al. (CMS Collaboration), Phys. Lett. B
793, 520 (2019)                                                                                    
\bibitem{GG1} Gayatri Ghosh, Nucl.Phys.B 979 (2022) 115759, arXiv: 2106.12503 [hep-ph                                                                                ]
\bibitem{GG2} Gayatri Ghosh, Indian J.Phys. (2024), arXiv: 2307.09948 


 \end{thebibliography}
	                                                                                                \end{document}